\documentclass{article}
\usepackage[dvips]{graphicx}

\author{N. Redington
\\ Net Advance of Physics
\\redingtn@mit.edu}
\title{On the Apparent Superluminal Motion of a Damped Gaussian Pulse}
\date{}

\begin{document}
\maketitle

\begin{quote}
Alicki has demonstrated that a travelling Gaussian pulse subject 
to damping is indistinguishable from 
 an undamped pulse moving with greater speed; such an effect could
create the illusion of a pulse moving faster than light. In this note, an
alternative derivation of  the same result is presented. However, it
is unlikely that this particular illusion could explain the superluminal
neutrino-velocities reported by OPERA. 
\end{quote}

\bigskip

The recent claim that the measured velocity of mu neutrinos exceeds that of light [1]  
has rekindled interest in the old subject of illusory superluminal motion. 
Of the various proposals advanced to explain the OPERA results without
jeopardy to established physics [2], that of Alicki [3] is among the most intriguing. 
Alicki argues that a stationary observer cannot tell a moving 
normal distribution subject to global exponential damping from
an undamped normal distribution with different characteristic parameters, 
including higher (even apparently superluminal) velocity.
This result, which may be of some importance in optics and other fields
unrelated to neutrino physics, is re-derived in the present note by very elementary 
means.  Unfortunately (or otherwise, if one wishes to break the light barrier
in fact rather than appearance), such an effect seems unlikely to be the solution to the
OPERA puzzle: the CNGS-beam pulse lacks the ``fat tail'' needed for the
argument to apply. 

To derive Alicki's result, it is first helpful to consider the case of a Gaussian
pulse with no damping at all. Let $\rho (x,t)$ denote the intensity, assumed non-negative, of
a signal moving at constant speed $v>0$ in one dimension without dispersion 
or distortion toward an observer at some fixed location $x = L$. By hypothesis,
observer and source are in a state of mutual rest. Let the profile of the 
signal as initially transmitted
be Gaussian with maximum height $\rho_{0}$ and full width $\sigma$ at half
maximum. Since transmission is perfect, the observer at $L$ will see:
$$\rho (t) =  \rho_{0}  \exp -  \frac{(L - vt)^{2}}{2 \sigma ^{2}}     $$
The time derivative of $\rho$ is:
$$ \dot{\rho} = \rho \frac{v}{\sigma^{2}} (L - vt) $$
so the observer will notice a peak at time
$$t= L/v$$
Prior to that time, $\rho$ is increasing; afterwards, decreasing. Knowing $L$
and measuring $t$, the observer easily determines the velocity of the pulse
to be $v$, as expected.

Now introduce a simple damping:
\begin{equation}
\rho (t) =  (\rho_{0}  \exp -  \frac{(L - vt)^{2}}{2 \sigma^{2}})   \exp -\gamma t 
\end{equation}
where $\gamma$ is a positive constant. The time derivative of $\rho$ becomes:
$$ \dot{\rho} = \rho \frac{v}{\sigma^{2}} (L - vt - \delta L)  $$
where 
\begin{equation}
\delta L \equiv  \frac{\gamma \sigma^{2}}{v} \geq 0
\end{equation}
The observer will now notice a peak at time
$$t= T \equiv (L - \delta L)/v$$
Prior to that time, $\rho$ is increasing; afterwards, decreasing.    
It is therefore reasonable to define 
\begin{equation}
v' \equiv L/T = \frac{Lv}{L - \delta L}  
\end{equation}
Note that $v' \ge v$ provided  $0 \le \delta L \le L$.

To shew that $v'$ could in fact be interpreted as the velocity of a moving,
undamped Gaussian pulse, note that according to Eq. (2):
\begin{equation}
\sigma^{2} \gamma t =   vt \delta L
\end{equation}
Therefore, we may rewrite Eq. (1) in terms of $\delta L$ as:
\begin{equation}
 \rho (t) = \rho_{0} \exp - \frac {F}{2 \sigma^{2}}
\end{equation}
where
\begin{equation}
F \equiv L^{2} + (vt)^{2}  - 2 (L - \delta L) vt 
\end{equation}
Furthermore, define two new constants
\begin{equation}
\sigma' \equiv \frac{\sigma}{1-(\delta L / L)}
\end{equation}   
and
\begin{equation}
\rho_{0}' \equiv \rho_{0} \exp - \frac{\delta L}{2 \sigma^{2}} (2L - \delta L)
\end{equation}
With the help of Eqs. (2), (3), (6), (7), and (8), Eq. (5) becomes:
\begin{equation}
\rho (t) =  \rho_{0}'  \exp -   \frac{(L - v't)^{2}}{2 \sigma'^{2}} 
\end{equation}
quod erat demonstrandum.  

Less formally, the Alicki effect can be understood in terms of the damping 
acting on the signal over time. At $t=0$, the value of $\rho$ in the neighbourhood
of $L$ is $\rho_{0} \exp - (L/ \sigma)^{2} / 2$.  A few moments later, the part of the
distribution originally just a bit to the left of $L$ will pass the observer; the damping
force will have acted upon it briefly, so its magnitude will be somewhat
diminished. By contrast, the regions of the distribution which at $t=0$ were far to
the left, though they may have had initially large values of
$\rho$, will arrive at the observer much later, and the exponential decay induced by
the damping force will have acted on them for a long time. Thus, the magnitude of
parts of a sufficiently wide Gaussian may be greatly diminished by even a small 
damping, and the peak of
the received signal will be shifted to the right relative to the peak in the signal as originally
sent.

The constraint $\delta L \le L$ for positive $v'$ can also be given a 
qualitative interpretation. Consider the three factors which define $\delta L$ 
according to Eq. (2). Should
the damping constant $\gamma$ be too large, or the initial velocity $v$ too small,
the magnitude of the signal reaching the observer will diminish from the very first
observation at $t=0$. The observer
might conclude from this that the pulse is moving to the left (in agreement with
Eq. (3) for this case), and that only its receding tail is being observed. The same is
true, if perhaps less self-evidently, for a very broad pulse: it rises too slowly to keep up with
the decay, and may also be construed as moving leftwards.   

Since both source and observer are stationary in a common frame of reference, the 
initial velocity $v$ may be given any value
up to $c$ without affecting the forgoing analysis. In the case $v=c$ with $\delta L \le L$, 
the apparent velocity seen by the observer will always be superluminal and may even
be infinite. However, no information can be transmitted faster than light in this way. 
Although the peak
of the distribution seen by the observer arrives much sooner than it would have in the 
absence of damping, its magnitude cannot exceed whatever value the undamped signal
would have had at that moment. Any message communicated by the arrival of the 
undamped peak could therefore have
been transmitted just as swiftly (or more so) by specifying a particular magnitude
of signal intensity as the communication threshold. 

In the real world, no signal is truly Gaussian: the intensity at the origin does not 
begin to rise above zero in the infinitely distant past. Rather, the sender of the 
signal presses a button at some time $t_{0} < 0$, generating output which rises
in approximately Gaussian fashion to a peak at $t=0$ and then dies away.
The first evidence of this signal -- the start of the long but finite rising tail -- reaches
the observer no sooner than some minimum time
$\tau  \equiv t_{0} + (L/v)$. Even though the peak of the
received signal may occur at a time less than $L/v$, no bit of information can
arrive before $t=\tau$, and thus the message ``the button has been pressed'' cannot
be transmitted with speed greater than $v$.

Alicki, of course, has never claimed that his effect can be used to send information
faster than light, or that it is more than an interesting illusion. He has, however,
suggested that it might explain  the OPERA superluminal neutrino data. This seems
to me to be unlikely. The pulses from the CNGS beam used in the experiment were 
not Gaussian: indeed, to a first approximation, they were  almost square. They entirely 
lacked the characteristic long tail of the normal distribution, the envelope above which
the Alicki peak cannot rise.  

By the ``arrival'' of the neutrino pulse, the OPERA group meant the arrival of the nearly vertical front, not that of the centre of the broad, relatively flat peak. It is true that a series of 
arguably bell-shaped  sub-pulses can be discerned in the flat region, and, in the 
second version of his paper [4], Alicki seems to treat these, rather than the whole pulse, 
as the Gaussian signals which have been shifted. This cannot, however, explain the 
superluminal motion of the square base-pulse above which the sub-peaks rise. 

Henri [5] has described a superluminal illusion involving logistic rather than Gaussian signals,  
clearly a better approximation to the actual shape of the CNGS-beam distribution-function. 
Broda [6] has suggested a general class of statistical models which may generate 
such illusions, given a particular method of maximum-likelihood estimation. The analysis 
of these more complicated proposals  is beyond the scope of the present note.  

Although it does not resolve the OPERA puzzle, the Alicki effect is an interesting
superluminal illusion, and one which experimentalists should be able to demonstrate
in the laboratory with acoustical or optical signals.

\bigskip
REFERENCES
\begin{small}
\begin{enumerate}
\item T. Adam et al., ``Measurement of the neutrino velocity with the OPERA detector in the CNGS beam''
arXiv:1109.4897
\item A partial list may be found at the Net Advance of Physics website:
\newline
http://web.mit.edu/redingtn/www/netadv/ftlNuSlow.html
\item R. Alicki, ``A possible statistical mechanism of anomalous neutrino velocity in OPERA experiment?'' (version 1) 
arXiv:1109.5727v1
\item R. Alicki, ``A possible statistical mechanism of anomalous neutrino velocity in OPERA experiment?'' (version 2)
arXiv:1109.5727v2
\item G. Henri, ``A simple explanation of OPERA results without strange physics''  
arXiv:1110.0239v1
\item B. Broda, ``An OPERA inspired classical model reproducing superluminal velocities''  
arXiv:1110.0644v3

\end{enumerate}
\end{small}

\end{document}